\documentclass[twocolumn,aps,showpacs,prx,amsmath,amssymb,floatfix,superscriptaddress]{revtex4-1}
\usepackage{color}
\usepackage{graphicx}
\usepackage{bm}
\usepackage{array}
\usepackage{float}
\usepackage{supertabular}
\usepackage{longtable}
\usepackage{mathrsfs}
\usepackage{txfonts}
\usepackage{wasysym}

\begin{document}

\title{Dynamics of noisy quantum systems in the Heisenberg picture: application to the stability of fractional charge}

\author{Armin Rahmani}

\affiliation{
Theoretical Division, T-4 and CNLS, Los Alamos National Laboratory, Los Alamos, New Mexico 87545, USA} 

\affiliation{Quantum Matter Institute, University of British Columbia, Vancouver, British Columbia, Canada V6T 1Z4}

\date{\today}

\begin{abstract}

Based on the Heisenberg-picture analog of the master equation, we develop a method for computing the exact time dependence of noise-averaged observables for general noninteracting fermionic systems with noisy fluctuations. Upon noise averaging, these fluctuations generate effective interactions, limiting analytical approaches. While the short-time dynamics can be studied with Langevin-type numerical simulations, the long-time limit is not amenable to such simulations. Our results provide access to this long-time limit. As a simple example, we examine the fate of the fractional charge in cold-atom emulations of polyacetylene after stochastic driving. We find that in a quantum quench to a fluctuating hopping Hamiltonian, the fractional charge remains robust for hopping between different sublattices, while it becomes unstable in the presence of noisy hopping on the same sublattice. 
\end{abstract}
 
\maketitle
\section{introduction}

Recent developments in atomic, molecular and optical physics have made it possible to create optical-lattice incarnations of important many-body Hamiltonians with a large degree of control and tunability~\citep{Bloch2008}. Such systems exhibit remarkable isolation from a thermal environment and can undergo coherent unitary evolution in experimentally accessible time scales. These developments have motivated numerous studies of the  nonequilibrium quantum dynamics of thermally isolated systems~\cite{Polkovnikov2011}. These systems are, nevertheless, vulnerable to noise-induced heating, which can originate from, e.g., amplitude fluctuations of the lasers forming the optical lattice. Due to the temporal fluctuations of the Hamiltonian, the time evolution is governed by a \textit{stochastic} Schr\"odinger equation (SSE).

Understanding the effects of noise in such systems is of paramount importance both in designing experiments and interpreting their results.  Several publications have investigated the heating dynamics due to such stochastic driving in specific systems such as harmonic traps~\cite{Gehm1998,Grotz2006}, Luttinger liquids~\cite{D'Alessio2013}, transverse-field Ising chain~\cite{Marino2012,Marino2014}, and the Bose-Hubbard model~\cite{Pichler2012,Pichler2013}, using various techniques and approximations (see also Ref.~\cite{Bunin2011} for generic results on energy fluctuations). However, the general problem remains unsolved analytically and the numerics are limited to short time scales.

The main challenge for computing \textit{noise-averaged} observables in systems governed by a many-body SSE stems from effective interactions that arise, even in systems that are completely quadratic for each realization of noise, upon integrating out the noise. This severely limits the application of field-theoretical techniques. Equivalently, one can formulate a Lindblad-type master equation for the noise-averaged density matrix. However, the effective noise-induced interactions make it necessary to work with an exponentially large Hilbert space. (For linear Lindblad operators, such interactions are not generated and the problem remains Gaussian~\cite{Prosen2008}.) For quadratic Lindblad operators, an algebraic structure has been found recently, which decouples the Lindblad equation~\cite{Znidaric2010,Eisler2011,Znidaric2011}, allowing for significant progress. Direct numerical approaches to the problem through Langevin-type simulations (assuming one can solve the dynamics numerically for each realization of noise) are  computationally demanding and can not access the long-time limit.


Building on the results of Ref.~\cite{Eisler2011}, we argue in this paper that a \textit{Heisenberg-picture analog of the master equation} gives the exact time dependence (including the long-time limit) of the noise-averaged expectation values of most physical observables (e.g., energy, fermionic Green's function, density-density correlations, etc.) in systems of lattice fermions with noisy quadratic Hamiltonians under the assumption is that the fluctuations have \textit{white-noise} character. This seemingly exponential improvement in the Heisenberg picture originates from the fact that the noise-averaged density matrix, which is central to the Schr\"odinger picture, can give the expectation value of \textit{any} operator. The simplification in the Heisenberg picture arises because our observables of interest have a few (e.g., 2 for Green's function and 4 for density-density correlation functions) creation/annihilation operators. This is reminiscent of the remarkable performance gains of the density-matrix-renormalization-group algorithm in the Heisenberg picture, where, instead of the entire state, only the observables of interest are kept~\cite{Hartmann2009}.

As an application of our formalism, we study a system relevant to the emulation of polyacetylene (the simplest system exhibiting topological properties and charge fractionalization) in an optical lattice~\cite{Atala2013}. The domain walls in this system bind a fractional charge. We compute the long-time limit of the charge-density profile after driving the system with noisy (and spatially disordered) hopping processes. We find that the fractional charge is robust against noisy fluctuations in the hopping processes as long as the fluctuations are limited to hopping between sites on different sublattices. The presence of noisy fluctuations in hopping on the same sublattice, on the other hand, globally distorts the charge-density profile, hindering the observation of the fractional charge.

The outline of this paper is as follows. In Sec.~\ref{sec:form}, we develop the formalism for solving the Heisenberg-picture analog of the master equation. Sec.~\ref{sec:frac} is focused on the fate of the fractional charge in a noisy dimerized chain. We close the paper in Sec.~\ref{sec:conc} with a brief summary. More details are given in three appendices.

\section{formalism}
\label{sec:form}

\subsection{General setup}
We begin by discussing the general formulation of the problem. Consider a quantum quench described by the following time-dependent Hamiltonian:
\begin{eqnarray}\label{eq:hamil}
H(t)&=&H_0,\quad t<0,\\
H(t)&=&H_1+\sum_i \alpha_i(t) V_i,\quad t>0,
\end{eqnarray}
where all the time dependence for $t>0$ is in the form of white noise $\alpha_i(t)$ with zero mean and second moment
 \begin{equation}\label{eq:white}
\overline{\alpha_i(t)\alpha_j(t')}=\delta_{ij}W_i^2 \delta(t-t').
\end{equation}
The system then evolves with an SSE
\begin{equation}\label{eq:sto}
\partial_t \rho(t) =-i\left[H_1+\sum_i\alpha_i(t)V_i, \rho(t)\right],
\end{equation}
where $\rho(t)$ is the density matrix for one realization of noise.
The above SSE is interpreted in the Stratonovich sense as we are dealing with continuous processes. It simply describes an ensemble of quantum evolutions. (Throughout the paper $\hbar$ is set to unity.) 

Given an initial density matrix $\rho(0)$ at $t=0$ (the same for all realizations of noise), the goal is to find $\overline{ \langle O(t)\rangle}={\rm tr}\left[\overline{\rho(t)}O\right]$, the noise-averaged expectation value of an operator $O$ after the system has evolved for a time $t$ with the SSE~\eqref{eq:sto} (the overline indicates noise averaging). Each realization of noise evolves the system deterministically, resulting in a unique quantum expectation value $\langle O(t)\rangle$. These are then averaged over all trajectories with their corresponding probabilities.

It is known in the theory of open quantum systems~\cite{Breuer2002,Barchielli2009} that certain Lindblad-type master equations can be ``unravelled'' into a set of quantum trajectories described by an SSE~\cite{Caves1987,Diosi2000}. Such mapping then allows for tacking the master equation numerically through Langevin-type simulations of the SSE. Such numerical simulations are limited to short time scales, but have proved useful in studying otherwise intractable master equations. Our interest here is in thermally isolated driven systems, where the SSE is the starting point. It may thus appear that the connection to a master equation is not very useful. However, a \textit{Heisenberg-picture} analog of the master equation can yield exact results for our systems of lattice fermions. In addition to driven systems, our results can also be applied to open systems, which can be unravelled into the types of SSE considered in this work.

 The master equation corresponding to SSE~\eqref{eq:sto} is $
{d \over dt }\overline{\rho(t)}=-i\left[H_1, \overline{\rho(t)}\right]+{1\over 2}\sum_j{W_j^2}\left[[V_j,\overline{\rho(t)}],V_j\right]$, which implies the following equation of motion for $\overline{O(t)}$:
\begin{equation}\label{eq:eom}
{d \over dt }\overline{O(t)}=i\left[H_1, \overline{O(t)}\right]+{1\over 2}\sum_j{W_j^2}\left[[V_j,\overline{O(t)}],V_j\right],
\end{equation}
where $\overline{O(t)}=\overline{U^\dagger(t)OU(t)}$ is the noise-averaged Heisenberg-picture operator at time $t$ (see Appendix.~\ref{app:a} for an elementary derivation). Here $U(t)$ is the evolution operator that depends on the trajectory.

We focus on systems of lattice fermions. We assume we have lattice with $L$ sites and represent quadratic operators
\begin{equation}
P=\sum_{i,j}{\mathscr P}_{ij}c^\dagger_i c_j,
\end{equation}
 where $c_i$ is the annihilation operator for a fermion on site $i$, by $L\times L$  ($i,j=1\cdots L$) matrices $\mathscr P$ through the shorthand notation
\begin{equation}\label{eq:gamma}
P=\Gamma({\mathscr P}).
\end{equation}
Higher-order operators such as $\sum_{ijkl}{\mathscr R}_{ijkl}c^\dagger_i c_jc^\dagger_k c_l$ can be represented by higher-rank tensors $\mathscr R$.

\subsection{Simplifications in the Heisenberg picture}

As mentioned in the introduction, we assume that the Hamiltonian is \textit{quadratic}. Even for a fully quadratic system, the noise-averaged effective theory is interacting (see, e.g, Refs.~\cite{DallaTorre2010,Marino2012,Wilson2012}). For simplicity, hereafter, we focus on the case with only one such quadratic fluctuating term 
\begin{equation}V=\Gamma({\mathscr V})
\end {equation}
but the generalization to more quadratic noise terms is straightforward. 
 For the master equation governing the \textit{noise-averaged} density matrix, the Gaussian ansatz, which is characteristic of noninteracting systems, breaks down. To see this explicitly, we consider the evolution of $\overline{\rho(t)}$ with the master equation with an initial density matrix that can be written as 
 \begin{equation}\rho(0)=e^{\Gamma[{\mathscr S}(0)]}
 \end {equation}
  for an $L\times L$ matrix ${\mathscr S}(0)$ (corresponding to natural initial states such as a thermal state with respect to a quadratic $H_0$). While the density matrix retains the form above [for a time-dependent ${\mathscr S}(t)$] for each realization of noise, the noise-averaged density matrix will not. This follows from inserting the ansatz $e^{\Gamma[{\mathscr S}(t)]}$ in the master equation: the double commutator gives a quartic form times $e^{\Gamma[{\mathscr S}(t)]}$, while all other terms produce a quadratic form times $e^{\Gamma[{\mathscr S}(t)]}$ (see Appendix.~\ref{app:b} for details). Consequently, the above ansatz can not satisfy the master equation and the Wick's theorem (which follows from such Gaussian density matrices) and the free-fermion picture do not survive the noise averaging.

Nevertheless, if we work in the Heisenberg picture with an operator $O$ that is a product of a finite number of creation/annihilation operators, as a consequence of the identity~\cite{Klich2003,*Klich2003'} 
\begin{equation}\label{eq:rel}
\left[\Gamma({\mathscr A}),\Gamma({\mathscr B})\right]=\Gamma(\left[{\mathscr A},{\mathscr B}\right]),
\end{equation}
the double commutator in Eq.~\eqref{eq:eom} does not generate higher-order terms for any quadratic $V$. For a quadratic operator $O$ this follows directly from Eq.~\eqref{eq:rel} and for higher-order operators from a combination of Eq.~\eqref{eq:rel} and the operator identity $[AB,C]=A[B,C]+[A,C]B$. For example, a quartic operator can be expanded into products of quadratic operators (see Appendix.~\ref{app:c} for details).

The above discussion indicates that for a quadratic operator $O=\Gamma({\mathscr O})$, the ansatz 
\begin{equation}
\overline{O(t)}=\Gamma({\mathscr O(t)})
\end{equation}
satisfies the equation of motion.
We can cast the master equation into a simpler form by representing the $L\times L$ matrix ${\mathscr O}(t)$ with a vector $| {\mathscr O}(t)\rangle$ of length $L^2$:
\begin{equation}\label{eq:ghj}
{d\over dt}| {\mathscr O}(t)\rangle={\cal K }|{\mathscr O}(t)\rangle,
\end{equation} 
for an $L^2\times L^2$  matrix 
\begin{equation}\label{eq:K}
\begin{split}
{\cal K }_{\alpha\beta,\eta\gamma}=&i\left(\mathscr{H}_{1\alpha \eta}\delta_{\beta \gamma}-\mathscr{H}_{1\gamma \beta}\delta_{\alpha \eta}\right)\\
&+{1\over 2}{W^2}\left(2 {\mathscr V}_{\alpha \eta}  {\mathscr V}_{ \gamma\beta} -\delta_{\alpha \eta}  {\mathscr V}^2_{ \gamma\beta}-\delta_{ \gamma\beta} {\mathscr V}^2_{\alpha \eta}\right),
\end{split}
\end{equation} 
with $H_1=\Gamma({\mathscr H}_1)$. Eq.~\eqref{eq:ghj} has a simple formal solution
\begin{equation}
| {\mathscr O}(t)\rangle=e^{{\cal K}t} |\ {\mathscr O}(0)\rangle.
\end{equation} 

\subsection{Generic steady state for $H_1\neq 0$}
\label{C0}
At this point, the time dependence of $\overline{\langle O(t)\rangle}$ can be readily obtained by exponentiating the $L^2\times L^2$ matrix $\cal K$ as follows. We diagonalize $\cal K$ as
\begin{equation}
{\cal K}={\cal U}{\cal D} {\cal U}^{-1} ,
\end{equation}
where ${\cal D}={\rm diag}(d_1,d_2,\dots d_{L^2})$. We can then write $e^{{\cal K}t}={\cal U} {\rm diag}(e^{d_1t},e^{d_2t},\dots e^{d_{L^2}}) {\cal U}^{-1}$.

As the matrix $\cal K$ is not generically Hermitian, it can have complex eigenvalues. On physical grounds, we do not expect $\cal K$ to have any eigenvalues with a positive real part as these would lead to the divergence of $|\ {\mathscr O}(t)\rangle$ in the limit of $t \to \infty$. Moreover, $e^{d_jt}$ for all eigenvalues $d_j$ with a negative real part decays to zero in the long-time limit. Therefore, the limit of $t\to\infty$ is dominated by eigenvalues $d_j$ with ${\rm Re}(d_j)=0$. Here we show that the matrix $\cal K$ in Eq.~\eqref{eq:K} has at least one eigenvector with a vanishing eigenvalue. The eigenvector (of length $L^2$) with $d_j=0$ corresponds to the $L\times L$ identity matrix $\openone_{\eta\gamma}=\delta_{\eta\gamma}$: one can readily verify that
\begin{eqnarray}
& &\sum_{\eta\gamma}\left(\mathscr{H}_{1\alpha \eta}\delta_{\beta \gamma}-\mathscr{H}_{1\gamma \beta}\delta_{\alpha \eta}\right)\delta_{\eta\gamma}=0,\\
& &\sum_{\eta\gamma}\left(2 {\mathscr V}_{\alpha \eta}  {\mathscr V}_{ \gamma\beta} -\delta_{\alpha \eta}  {\mathscr V}^2_{ \gamma\beta}-\delta_{ \gamma\beta} {\mathscr V}^2_{\alpha \eta}\right)\delta_{\eta\gamma}=0.
\end{eqnarray}

If $\openone$ is the only eigenvector of $\cal K$ with ${\rm Re}(d_j)=0$, then the  $t\to\infty$ fate of $\overline{\langle O(t)\rangle}$ for quadratic operators is simple. Unlike the case of $H_1=0$ discussed in Sec.~\ref{C}, for a generic $H_1\neq 0$, there is no reason for the vanishing eigenvalue to be degenerate (having another eigenvalue with a vanishing real part would be a nongeneric accidental occurrence for $H_1\neq 0$, which we do not consider here). The long-time limit of the evolution operator $e^{{\cal K}t}={\cal U}e^{{\cal D}t} {\cal U}^{-1}$ can the be evaluated as follows.  In the $t\to\infty$ limit, all the elements of the diagonal matrix $e^{{\cal D}t} $ decay to zero except for the one corresponding to $d_j=0$. Then the only column of $\cal U $ that survives in the evolution operator is the one corresponding eigenvector $|\openone\rangle$. As the matrix $\cal K$ is not generically Hermitian, ${\cal U}^{-1}\neq {\cal U}^\dagger$ and generic rows of ${\cal U}^{-1}$ are not the Hermitian conjugates of the eigenvectors of $\cal K$. However, the row in  ${\cal U}^{-1}$  that corresponds to $d_j=0$ is indeed equal to $\langle \openone |$ as $\openone$ is both a right and a left eigenvector of $\cal K$ with a vanishing eigenvalue. Thus, for the generic case with only one eigenvector of $\cal K$ with ${\rm Re}(d_j)=0$, we can finally write
\begin{equation}
\lim_{t\to\infty}e^{{\cal K}t}=|\openone\rangle\langle \openone |,
\end{equation}
which projects $L\times L$ matrices onto ${1\over L}\openone$ (note that we assume the state $|\openone\rangle$ is normalized).

We now consider the fate of the two-point functions $O=c^\dagger_ic_j$ under this generic drive. For $i\neq j$, the corresponding matrix $\mathscr O_{\eta\gamma}=\delta_{i\eta}\delta_{j\gamma}$ has no overlap with $\openone_{\eta\gamma}=\delta_{\eta\gamma}$ and the noise-averaged Green's function decays to zero. For $i=j$, all $\mathscr O$ have the same overlap with $\openone$ (corresponding to the total density) and the time evolution results in a uniform noise-averaged density. In the following section we focus on a quench with $H_1=0$ and demonstrate that $\cal K$ will generically have $L$ vanishing eigenvalues in that case, leading to a different generic behavior.

\subsection{Explicit solution and nontrivial steady states for $H_1=0$} 

\label{C}

 In this section, we focus on a quantum quench, where $H_1=0$ and $\cal K$ is, consequently, Hermitian. It turns out that in this case the eigenvalues and eigenvectors of $\cal K$ in Eq.~\eqref{eq:K} are related to those of the $L\times L$ matrix $\mathscr V$ and we only need to diagonalize this smaller matrix. We write
 \begin{equation}
 {\mathscr V}={\mathscr U}{\mathscr D}{\mathscr U}^\dagger,
 \end{equation}
for a diagonal matrix ${\mathscr D}$ and a unitary ${\mathscr U}$. Upon inserting the above expression for $\mathscr V $ into Eq.~\eqref{eq:K} (for ${\mathscr H}_1=0$) and writing the Kronecker $\delta_{\alpha \eta}=\sum_\sigma {\mathscr U}_{\alpha \sigma} {\mathscr U}^\dagger_{ \sigma \eta}$ (and similarly for $\delta_{ \gamma\beta} $), we obtain
\begin{equation} {\cal K }_{\alpha\beta,\eta\gamma}=-{1\over 2}{W^2}\sum_{\sigma \lambda}{\mathscr U}_{\alpha \sigma} {\mathscr U}^\dagger_{ \lambda \beta}
\left({\mathscr D}_{\sigma \sigma }-{\mathscr D}_{\lambda \lambda}\right)^2
{\mathscr U}^\dagger_{ \sigma \eta}{\mathscr U}_{ \gamma \lambda}.
\end{equation}
 The matrix $e^{{\cal K}t} $ can now be simply written in terms of the above diagonalized form, which leads to
\begin{equation}\label{eq:exact}
  {\mathscr O}_{\alpha \beta}(t)=\sum_{\sigma \lambda \eta \gamma}{\mathscr U}_{\alpha \sigma} {\mathscr U}^\dagger_{ \lambda \beta}
e^{-{W^2\over 2}\left({\mathscr D}_{\sigma \sigma }-{\mathscr D}_{\lambda \lambda}\right)^2 t}
{\mathscr U}^\dagger_{ \sigma \eta}{\mathscr U}_{ \gamma \lambda}  {\mathscr O}_{\eta \gamma},
\end{equation}
where we have used the boundary condition $  {\mathscr O}(0)={\mathscr O}$.
Note that none of the eigenvalues $-{W^2\over 2}\left({\mathscr D}_{\sigma \sigma }-{\mathscr D}_{\lambda \lambda}\right)^2$ of $\cal K$ are positive so $  {\mathscr O}(t)$ does not diverge in the  $t\rightarrow \infty$ limit.
Assuming there are no degeneracies in the spectrum of $\mathscr V$, the matrix $\cal K$ has $L$ vanishing eigenvalues for $\sigma =\lambda$ (corresponding to eigenvalues of $e^{{\cal K}t}$, which do not decay exponentially). The limit of long times can then be easily accessed by setting these decaying eigenvalues of $e^{{\cal K}t} $ to zero. We then obtain
\begin{equation}\label{eq:long}
  {\mathscr O}_{\alpha \beta}(t\rightarrow \infty)=\sum_{\sigma \eta \gamma} {\mathscr U}_{\alpha\sigma }
 {\mathscr U}_{\sigma \beta}^\dagger
  {\mathscr U}_{\sigma \eta}^\dagger
  {\mathscr U}_{\gamma\sigma } 
 {\mathscr O}_{\eta \gamma}.
\end{equation}

We can now use Eq.~\eqref{eq:exact} above and its $t\rightarrow \infty$ limit~\eqref{eq:long} to write the exact time dependence and the long-time limit of the noise-averaged expectation value of an operator $O=\Gamma({\mathscr O})$ by computing the expectation values of these explicit operators with the initial state. For example, if $H_0$ is quadratic and we are initially in its ground state, we can write
\begin{equation}\label{eq:final}
\overline{\langle O(t)\rangle}=\sum'_\alpha \left({\mathscr U}_0^\dagger   {\mathscr O}(t){\mathscr U}_0\right)_{\alpha \alpha},
\end{equation} 
where the ``prime'' symbol indicates summing over the initially occupied single-particle levels in the ground state of $H_0=\Gamma({\mathscr H}_0)$, and the unitary matrix ${\mathscr U}_0$ diagonalizes ${\mathscr H}_0$. For the Green's function $\langle c^\dagger_i c_j \rangle$, we have ${\mathscr O}_{\eta \gamma}=\delta_{i\eta}\delta_{j\gamma}$ and a special case $i=j$ gives the local density.

\section{The stability of fractional charge}\label{sec:frac}

\begin{figure}[top]
 \includegraphics[width =6 cm]{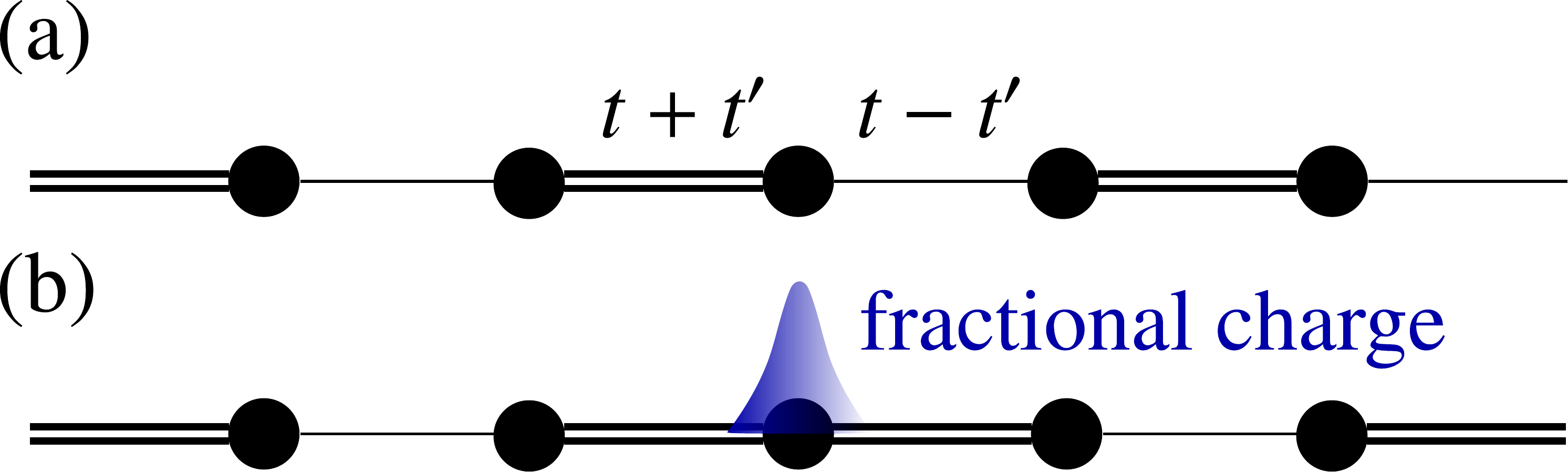}
 \caption{(a) The dimerized hopping model of Eq.~\eqref{eq:poly}. (b) A domain wall configuration with fractional charge.}
 \label{fig:2}
\end{figure}

We now focus on a simple model (relevant to polyacetylene), which exhibits an interesting topological property, namely, a fractional charge bound to domain walls~\cite{Jackiw1976,Su1979}. This model (without the domain wall at this point) has been recently implemented in optical lattices~\cite{Atala2013}. As shown in Fig.~\ref{fig:2}(b), we consider spinless fermions hopping on a one-dimensional lattice, with the hopping amplitude modulated as follows:
\begin{equation}\label{eq:poly}
H_0=\sum_x \left[t+(-1)^xt'\right]\left(c_x^\dagger c_{x+1}+c_{x+1}^\dagger c_x\right).
\end{equation}  
For $t'=0$, the low-energy effective description of the system consists of linearly dispersing massless Dirac fermions. The term proportional to $t'$ opens a gap by making the fermions massive (the sign of the mass is the same as the sign of $t'$). It is well-known that the two signs of mass correspond to two topologically distinct phases, and, thus, a domain wall [as shown in Fig.~\ref{fig:2}(b)] entails a change in the mass sign, resulting in a zero mode and fractional charge.

\begin{figure}
 \includegraphics[width =7 cm]{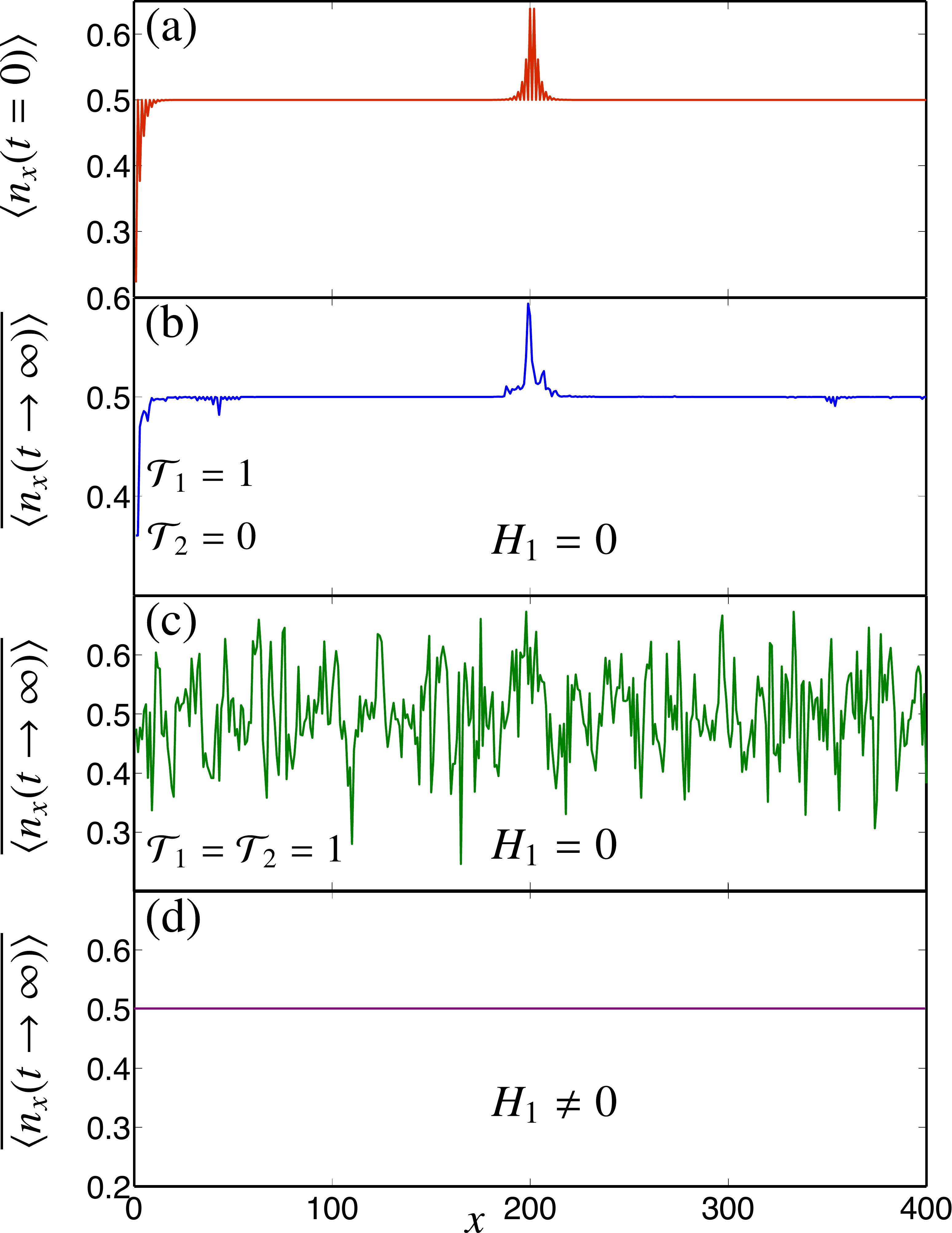}
 \caption{(a) the charge density profile in the ground state of Hamiltonian~\eqref{eq:poly} with a domain wall as shown in Fig.~\ref{fig:2}(b). (b) The noise-averaged charge density profile in the limit of $t\rightarrow \infty$ for noisy nearest-neighbor perturbations for $H_1=0$. (c) The noise-averaged charge density profile in the limit of $t\rightarrow \infty$ for noisy perturbations up to second neighbor for $H_1=0$. (d) The same long-time-limit density for the generic case of $H_1\neq 0$ discussed in Sec.~\ref{C0}.}
 \label{fig:3}
\end{figure}

We focus on a quantum quench with $H_1=0$ as in Sec.~\ref{C}. For the fluctuating component $V$, we consider local random hopping for up to second neighbor 
\begin{equation}
V=\sum_x\left[t^1_x\left(c_x^\dagger c_{x+1}+c_{x+1}^\dagger c_x\right)+t^2_x\left(c_x^\dagger c_{x+2}+c_{x+2}^\dagger c_x\right)\right],
\end{equation}
where $t^b_x$ is drawn from a uniform distribution $\left[-{{\cal T}_b\over 2}, {{\cal T}_b\over 2}\right]$. As for the operator $O$, we are interested in the local charge density $n_x=c^{\dagger}_x c_x$ at site $x$. We choose the ground state of $H_0$ as the initial state, which corresponds to half filling (with $L/2$ particles in a system of $L$ sites, where $L$ is even). The total number of fermions is then a constant of motion. In Fig.~\ref{fig:3}(a), we show the numerically computed charge density profile for a system of $L=400$ sites with a domain wall in the middle. We set the hopping $t$ to unity and express all other hoping amplitudes as dimensionless numbers in units of $t$. In Fig.~\ref{fig:3}(a), we used $t'=0.2$ but the actual value is inessential. As expected, the fractional charge bound to the domain wall is equal to $1/2$. Away from the domain wall and boundaries, the charge density is uniform $\langle n_x\rangle=1/2$.

In Figs.~\ref{fig:3}(b) and \ref{fig:3}(c), we show the long-time limit of the noise-averaged charge density profile $\overline {\langle n_x(t\rightarrow \infty) \rangle}$ computed using Eqs.~\eqref{eq:long} and \eqref{eq:final}, respectively for the case of random nearest neighbor hopping ${\cal T}_1=1$ and ${\cal T}_2=0$ and random first and second neighbor hopping ${\cal T}_1={\cal T}_2=1$. Notice that the overall scale of $V$ and the strength of noise $W$ can just affect how fast the $t \rightarrow \infty$ limit is reached but do not change the steady-state profile, which only depends on eigenfunctions of $\mathscr V$. Interestingly, the fractional charge remains robust for fluctuations in the nearest-neighbor hopping but becomes unstable as soon as second-neighbor hopping processes are included. 

This robustness is not a consequence of noise averaging and appears for each realization of noise. A hand-waving argument for the marked difference between Figs.~\ref{fig:3}(b) and \ref{fig:3}(c) can by made by considering $\overline {\langle n_x\rangle}$ in regions far away from the domain wall or the boundaries (they should behave in a similar way to $\overline {\langle n_x\rangle}$ in an infinite system without a domain wall at half filling). We consider two sublattices
\begin{equation}
a_i\equiv c_{2i},\quad b_i=c_{2i+1}.
\end{equation}
The Hamiltonians $H_0$ and $V$ for ${\cal T}_2=0$ have only terms that connect two different sublattices and can be written as $\sum_{ij} \left(t_{ij} a^\dagger_ib_j+{\rm H.c.}\right)$. However, $V$ for ${\cal T}_2\neq0$ contains terms of the form $a^\dagger_ia_j$ and $b^\dagger_ib_j$. We now consider the transformation 
\begin{equation}\label{eq:symm}
a_i\to a^\dagger_i, \quad b_i\to -b^\dagger_i, 
\end{equation}
which for real $t_{ij}$ (that could be time-dependent) maps $\sum_{ij} \left(t_{ij} a^\dagger_ib_j+{\rm H.c.}\right)$ onto itself. Note that we do not have this Hamiltonian symmetry in the presence of terms like $a^\dagger_ia_j$ and $b^\dagger_ib_j$, which flip sign for $i\neq j$ under the transformation~\eqref{eq:symm}. Also due to the particle-hole nature of the transformation, it only applies to half-filled systems. Importantly, $n_i\to 1-n_i$ under~\eqref{eq:symm}.  This implies that, if we do not have terms of type $a^\dagger_ia_j$ and $b^\dagger_ib_j$ in the time-dependent Hamiltonian, in the bulk of the system, the expectation value of $n_i$ must remain equal to $1/2$ during the evolution. This explains why a globally disordered density profile can only appear if sites on the same sublattice are connected.

It may appear that the symmetry argument above is inconsistent with the fact that there is fractional charge in the ground state of $H_0$ in Eq.~\eqref{eq:poly} in the first place. How can we have sites with $\langle n_i\rangle\neq 1/2$ if a Hamiltonian symmetry maps $n_i$ to $1- n_i$? In the presence of the domain wall, the ground state is degenerate (there are two single-particle zero modes in the spectrum and only one is occupied at half filling). Therefore, unlike the case of a unique ground state, the ground state is not invariant under the symmetry above and $\langle n_i\rangle$ does not need to equal $1-\langle n_i\rangle$. However, we expect this symmetry argument to capture the essential difference between Figs.~\ref{fig:3}(b) and \ref{fig:3}(c) when applied to the bulk of the system. As shown in Fig.~\ref{fig:3}(b), the generic case of $H_1\neq 0$ leads to a uniform density (see Sec.~\ref{C0} for a discussion).

%
%

\section {summary}\label{sec:conc}
In summary, we studied, for systems of quadratic lattice fermions, the dynamical effects of stochastic Hamiltonian noise on the variations of observables. We argued that a Heisenberg-picture approach to problem is more powerful than a Schr\"odinger-picture master equation. Despite the fact that the effective noise-averaged theory is a complex interacting one, we were able to calculate the exact time dependence using free-fermion techniques, providing access to the long-time limit.

We applied our formalism to the stability of fractional charge in a one-dimensional dimerized lattice with a domain wall, which is a promising candidate for realizing the phenomenon of fractionalization in optical lattices. We found an instability in the charge density profile in the presence of fluctuating second-nearest neighbor hopping processes, while the fractional charge remains robust after the quench to a fluctuating Hamiltonian if the fluctuations are limited to nearest-neighbor hopping. Application to the stability of fractional charge in two-dimensional quadratic systems such as graphene~\cite{Hou2007,Seradjeh2008} and frustrated itinerant magnets~\cite{Rahmani2013b} calls for future investigations. Also, extending our results to colored noise as well as to fluctuations around time-dependent Hamiltonians (for studying, e.g., the robustness of quantum annealing and optimal-control protocols~\cite{Chen2010,Doria2011,Rahmani2011,Choi2012,Ruschhaupt2012, Rahmani2013a}) are of considerable interest.

\acknowledgements
I am grateful to Cristian Batista, Adolfo del Campo, Claudio Chamon, Luca D'Alessio, Eugene Demler, Chang-Yu Hou, Ivar Martin, Anatoli Polkovnikov, and Kun Yang for helpful comments and discussions. This work was supported by the U.S. DOE under LANL/LDRD program, NSERC, and Max Planck-UBC Centre for Quantum Materials.

\appendix{}
\section{Derivation of the Heisenberg equation}\label{app:a}

\begin{figure}
 \includegraphics[width =8 cm]{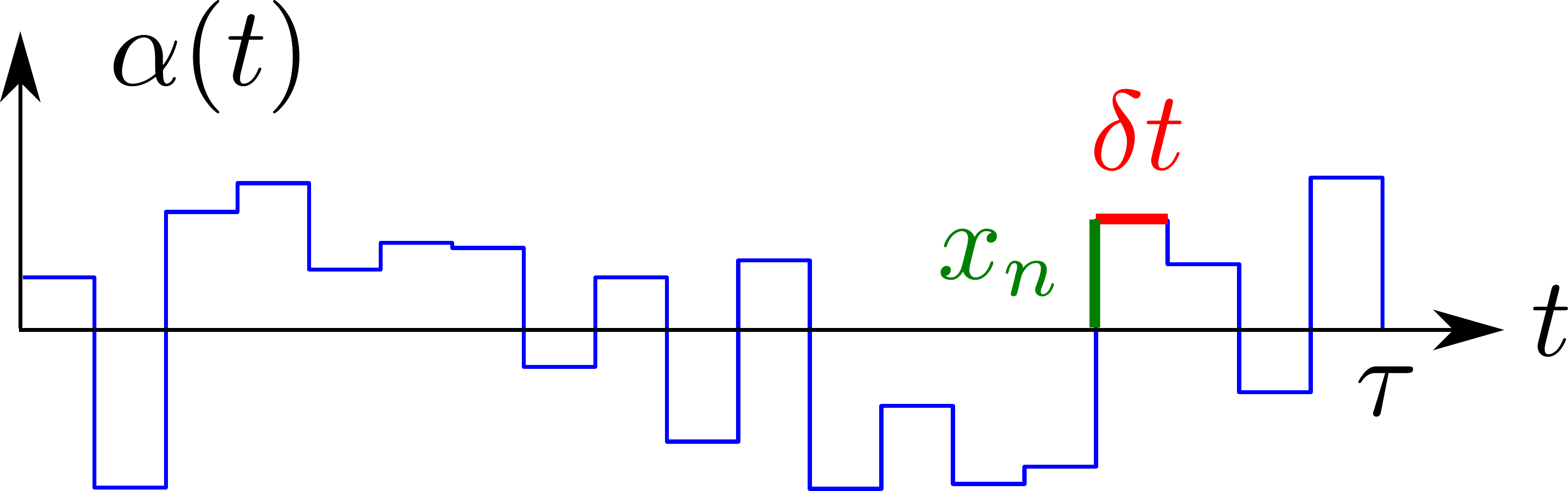}
 \caption{Discretized version of a white-noise protocol. Each $x_n$ is drawn from a uniform distribution  $[-\sqrt{3}{W\over\sqrt{\delta t}},\sqrt{3}{W\over\sqrt{\delta t}}]$ and $\delta t=\tau/N\rightarrow 0$.}
 \label{fig:1}
\end{figure}

Here, we present an elementary derivation of the Heisenberg-picture analog of the master equation. We can analyze the problem by approximating the white noise with an ensemble of discrete piece-wise constant protocols shown in Fig.~\ref{fig:1}. The total time $\tau$ for the stochastic evolution is divided into $N$ intervals of length $\delta t=\tau/N$ with $N \rightarrow \infty$ [$\alpha_i(t)=x^i_n$ is assumed constant over interval $n$] and each $x^i_n$ is drawn from a uniform distribution $[-\sqrt{3}{W_i\over\sqrt{\delta t}},\sqrt{3}{W_i\over\sqrt{\delta t}}]$ such that $\overline{x^i_n}=0$ and $\overline{x^i_n x^i_m}={W_i^2 \over \delta{t}}\delta_{mn}$. It is convenient to further define
\begin{equation}
w^i_n\equiv x^i_n \sqrt{\delta t}, \qquad \overline{w^i_n}=0,\qquad \overline{w^i_n w^i_m}=W_i^2 \delta_{mn}.
\end{equation}

The noise-averaged expectation value of a time-independent operator $O$ at time $\tau$ is then given by
\begin{equation}\label{eq:expect}
\langle \overline{O(\tau)} \rangle={\rm tr}\left[\rho(0)\overline{U^\dagger(\tau) O U(\tau) }\right],
\end{equation}
 where the evolution operator $U(\tau)$ for a fixed  $\{w^j_n\}$ realization of noise can be written as $U(\tau)=e^{-iH_1\delta t -i\sum_jV_j w^j_N \sqrt{\delta t}}\dots e^{-iH_1\delta t -i\sum_jV_j w^j_1 \sqrt{\delta t}}$, with $e^{-iH_1\delta t -i\sum_jV_j w^j_n \sqrt{\delta t}}$ approximately given by
\begin{equation}\label{eq:expan}
1-i\sum_jV_j w^j_n \sqrt{\delta t}-iH_1\delta t-{1\over 2}\sum_{jk}V_j V_kw^j_nw^k_n\delta t,
\end{equation}
in the limit of $N\rightarrow \infty$ ($\delta t\rightarrow 0$).
As the $w^j_n$ are uncorrelated for different $n$, we can compute $\overline{O(t+\delta t)}$ by acting on $\overline{O(t)}$ with Eq.~\eqref{eq:expan} and its Hermitian conjugate, respectively from the left- and the right-hand side, and then performing a noise averaging over only one set of stochastic variables $w^j_1$.  We obtain three nonvanishing terms (to order $\delta t$), which involve an even number of the same stochastic variable (and thus survive the noise averaging): one from taking the two $iV w^j_1 \sqrt{\delta t}$ terms from the two sides and two others from taking one  ${1\over 2}V_j^2w^2_n\delta t$ from each of the two sides. The above argument leads to the following equation of motion for $\overline{O(t)}$:
\begin{equation}\label{eq:eom2}
{d \over dt }\overline{O(t)}=i\left[H_1, \overline{O(t)}\right]+{1\over 2}\sum_j{W_j^2}\left[[V_j,\overline{O(t)}],V_j\right].
\end{equation}

\section{Evolution of the density matrix}\label{app:b}
We consider the time evolution of the density matrix of form $\rho(t)=e^{\Gamma[{\mathscr S}(t)]}$, where ${\mathscr S}(t)$ is an $L\times L$ matrix, with a deterministic quadratic Hamiltonian $\Gamma[{\mathscr H}(t)]$:
\begin{equation}\label{eom0}
\dot{\rho} =-i\left[\Gamma({\mathscr H}),\rho\right],
\end{equation} 
 where the ``dot'' symbol represents the time derivative. Combining Eq.~\eqref{eq:gamma} with the Baker-Campbell-Hausdorff formula $e^X Ye^{-X}=Y+[X,Y]+{1\over 2!}[X,[X,Y]]+{1\over 3!}[X,[X,[X,Y]]]+\dots$, leads to the identity
\begin{equation}\label{baker}
e^{\Gamma({\mathscr X})}\Gamma({\mathscr Y})e^{-\Gamma({\mathscr X})}=\Gamma\left(e^{{\mathscr X}}{\mathscr Y}e^{-{\mathscr X}}\right).
\end{equation} 
Using the identity above, we can write
 \begin{equation}\label{eq:id9}
\left[\Gamma({\mathscr H}),e^{\Gamma({\mathscr S})}\right]=\Gamma({\mathscr H}-e^{\mathscr S} {\mathscr H} e^{-{\mathscr S}})e^{\Gamma({\mathscr S})}.
\end{equation}
 Also, the identity
\begin{equation}
{d\over dt}e^{X(t)}=\int_0^1 du e^{uX(t)}\dot{X}(t)e^{(1-u)X(t)},
\end{equation} 
where $u$ is a dummy integration variable, together with Eq.~\eqref{baker} implies that the ansatz $\rho(t)=e^{\Gamma[{\mathscr S}(t)]}$ satisfies the equation of motion~\eqref{eom0} if ${\mathscr H}-e^{\mathscr S} {\mathscr H} e^{-{\mathscr S}}=i\int_0^1 du e ^{u{\mathscr S}} \dot{{\mathscr S}} e^{-u{\mathscr S}}$.

 However, if we consider the noise-averaged density matrix $\overline{\rho(t)}$, we have to add a double commutator ${W^2\over 2}\left[\left[\Gamma({\mathscr V}), e^{\Gamma({\mathscr S})}\right],\Gamma({\mathscr V})\right] $ to the right-hand side of the equation of motion. Using Eq.~\eqref{eq:id9}, the double commutator gives a quartic operator times $e^{\Gamma[{\mathscr S}(t)]}$:
 \begin{widetext}
\begin{equation}
\left[\left[\Gamma({\mathscr V}), e^{\Gamma({\mathscr S})}\right],\Gamma({\mathscr V})\right] =\left[\Gamma({\mathscr V}-e^{\mathscr S} {\mathscr V} e^{-{\mathscr S}})\Gamma(e^{\mathscr S} {\mathscr V} e^{-{\mathscr S}})-\Gamma({\mathscr V})\Gamma({\mathscr V}-e^{\mathscr S} {\mathscr V} e^{-{\mathscr S}})\right]
e^{\Gamma({\mathscr S})},
\end{equation} 
\end{widetext}
while, as we saw above, all the other terms reduce to quadratic operators times $e^{\Gamma[{\mathscr S}(t)]}$, indicating that the ansatz $e^{\Gamma[{\mathscr S}(t)]}$ breaks down when applied to the noise-averaged density matrix. Here we are assuming that $
\mathscr S$ does not commute with $\mathscr V$ and ${\mathscr V}-e^{\mathscr S} {\mathscr V} e^{-{\mathscr S}}\neq 0$ (the dynamics is trivial if they commute). 

\section{Closure of algebra for more general operators}\label{app:c}
We first consider the case of quartic charge-conserving operators.
We start with the commutation relation
\begin{equation*}
\left[c^\dagger_i c_j,c^\dagger_k c_l\right]=\delta_{jk}c^\dagger_i c_l-\delta_{il}c^\dagger_k c_j,
\end{equation*} 
which gives
\begin{equation*}
\begin{split}
\left[c^\dagger_i c_jc^\dagger_k c_l,c^\dagger_p c_q \right]=&
\delta_{lp}c^\dagger_i c_jc^\dagger_k c_q
-\delta_{kq}c^\dagger_i c_jc^\dagger_p c_l\\
&+\delta_{jp}c^\dagger_i c_qc^\dagger_k c_l
-\delta_{iq}c^\dagger_p c_jc^\dagger_k c_l.
\end{split}
\end{equation*} 
The above expression then implies that for $R=\sum_{ijkl}{\mathscr R}_{ijkl}c^\dagger_i c_jc^\dagger_k c_l$, we have
\begin{equation*}
\left[R,\Gamma({\mathscr V})\right]=\sum_{ijkl}{\mathscr T}_{ijlk}c^\dagger_i c_jc^\dagger_k c_l,
\end{equation*} 
with 
\begin{equation*}
{\mathscr T}_{ijlk}\equiv\sum_\alpha\left({\mathscr R}_{ijk\alpha}{\mathscr V}_{\alpha l}-
{\mathscr R}_{ij\alpha l}{\mathscr V}_{k \alpha }
+{\mathscr R}_{i\alpha kl}{\mathscr V}_{ \alpha j}
-{\mathscr R}_{\alpha ikl}{\mathscr V}_{ i\alpha}
\right).
\end{equation*}
Similarly, the closure of algebra holds if we have pairing terms. If we define a more general quadratic form
 \begin{equation*}\Gamma_p({\mathscr A},{\mathscr B},{\mathscr C},{\mathscr D})\equiv\sum_{ij}{\mathscr A}_{ij}c^\dagger_ic_j+{\mathscr B}_{ij}c^\dagger_ic^\dagger_j+{\mathscr C}_{ij}c_ic_j+{\mathscr D}_{ij}c_ic^\dagger_j,
\end{equation*}
 for $L\times L$ matrices ${\mathscr A}\dots{\mathscr D} $, we can show that
 \begin{equation*}
 \left[\Gamma_p({\mathscr A},{\mathscr B},{\mathscr C},{\mathscr D}),\Gamma_p({\mathscr A'},{\mathscr B'},{\mathscr C'},{\mathscr D'})\right]=\Gamma_p({\mathscr A''},{\mathscr B''},{\mathscr C''},{\mathscr D''}),
\end{equation*}
where
\begin{eqnarray*}
{\mathscr A''}&=&[{\mathscr A},{\mathscr A'}]-[{\mathscr A},{\mathscr D'}^T]+{\mathscr B}^{AS}{\mathscr C'}-{\mathscr B'}^{AS}{\mathscr C},\\
{\mathscr B''}&=&{\mathscr A}{\mathscr B'}^{AS}-{\mathscr A'}{\mathscr B}^{AS}+{\mathscr B}^{AS}{\mathscr D'}-{\mathscr B'}^{AS}{\mathscr D},\\
{\mathscr C''}&=&{\mathscr C}^{AS}{\mathscr A'}-{\mathscr C'}^{AS}{\mathscr A}+{\mathscr D}{\mathscr C'}^{AS}-{\mathscr D'}{\mathscr C}^{AS},\\
{\mathscr D''}&=&[{\mathscr D},{\mathscr D'}]-[{\mathscr D},{\mathscr A'}^T]+{\mathscr C}{\mathscr B'}^{AS}-{\mathscr C'}{\mathscr B}^{AS},
\end{eqnarray*}
with ${\mathscr B}^{AS}\equiv {\mathscr B}-{\mathscr B}^{T}$, where $T$ denotes matrix transpose.
\bibliography{heating}{}

\begin{thebibliography}{35}%
\makeatletter
\providecommand \@ifxundefined [1]{%
 \@ifx{#1\undefined}
}%
\providecommand \@ifnum [1]{%
 \ifnum #1\expandafter \@firstoftwo
 \else \expandafter \@secondoftwo
 \fi
}%
\providecommand \@ifx [1]{%
 \ifx #1\expandafter \@firstoftwo
 \else \expandafter \@secondoftwo
 \fi
}%
\providecommand \natexlab [1]{#1}%
\providecommand \enquote  [1]{``#1''}%
\providecommand \bibnamefont  [1]{#1}%
\providecommand \bibfnamefont [1]{#1}%
\providecommand \citenamefont [1]{#1}%
\providecommand \href@noop [0]{\@secondoftwo}%
\providecommand \href [0]{\begingroup \@sanitize@url \@href}%
\providecommand \@href[1]{\@@startlink{#1}\@@href}%
\providecommand \@@href[1]{\endgroup#1\@@endlink}%
\providecommand \@sanitize@url [0]{\catcode `\\12\catcode `\$12\catcode
  `\&12\catcode `\#12\catcode `\^12\catcode `\_12\catcode `\%12\relax}%
\providecommand \@@startlink[1]{}%
\providecommand \@@endlink[0]{}%
\providecommand \url  [0]{\begingroup\@sanitize@url \@url }%
\providecommand \@url [1]{\endgroup\@href {#1}{\urlprefix }}%
\providecommand \urlprefix  [0]{URL }%
\providecommand \Eprint [0]{\href }%
\providecommand \doibase [0]{http://dx.doi.org/}%
\providecommand \selectlanguage [0]{\@gobble}%
\providecommand \bibinfo  [0]{\@secondoftwo}%
\providecommand \bibfield  [0]{\@secondoftwo}%
\providecommand \translation [1]{[#1]}%
\providecommand \BibitemOpen [0]{}%
\providecommand \bibitemStop [0]{}%
\providecommand \bibitemNoStop [0]{.\EOS\space}%
\providecommand \EOS [0]{\spacefactor3000\relax}%
\providecommand \BibitemShut  [1]{\csname bibitem#1\endcsname}%
\let\auto@bib@innerbib\@empty
\bibitem [{\citenamefont {Bloch}\ \emph {et~al.}(2008)\citenamefont {Bloch},
  \citenamefont {Dalibard},\ and\ \citenamefont {Zwerger}}]{Bloch2008}%
  \BibitemOpen
  \bibfield  {author} {\bibinfo {author} {\bibfnamefont {I.}~\bibnamefont
  {Bloch}}, \bibinfo {author} {\bibfnamefont {J.}~\bibnamefont {Dalibard}}, \
  and\ \bibinfo {author} {\bibfnamefont {W.}~\bibnamefont {Zwerger}},\
  }\href@noop {} {\bibfield  {journal} {\bibinfo  {journal} {Rev. Mod. Phys.}\
  }\textbf {\bibinfo {volume} {80}},\ \bibinfo {pages} {885} (\bibinfo {year}
  {2008})}\BibitemShut {NoStop}%
\bibitem [{\citenamefont {Polkovnikov}\ \emph {et~al.}(2011)\citenamefont
  {Polkovnikov}, \citenamefont {Sengupta}, \citenamefont {Silva},\ and\
  \citenamefont {Vengalattore}}]{Polkovnikov2011}%
  \BibitemOpen
  \bibfield  {author} {\bibinfo {author} {\bibfnamefont {A.}~\bibnamefont
  {Polkovnikov}}, \bibinfo {author} {\bibfnamefont {K.}~\bibnamefont
  {Sengupta}}, \bibinfo {author} {\bibfnamefont {A.}~\bibnamefont {Silva}}, \
  and\ \bibinfo {author} {\bibfnamefont {M.}~\bibnamefont {Vengalattore}},\
  }\href@noop {} {\bibfield  {journal} {\bibinfo  {journal} {Rev. Mod. Phys.}\
  }\textbf {\bibinfo {volume} {83}},\ \bibinfo {pages} {863} (\bibinfo {year}
  {2011})}\BibitemShut {NoStop}%
\bibitem [{\citenamefont {Gehm}\ \emph {et~al.}(1998)\citenamefont {Gehm},
  \citenamefont {O'Hara}, \citenamefont {Savard},\ and\ \citenamefont
  {Thomas}}]{Gehm1998}%
  \BibitemOpen
  \bibfield  {author} {\bibinfo {author} {\bibfnamefont {M.~E.}\ \bibnamefont
  {Gehm}}, \bibinfo {author} {\bibfnamefont {K.~M.}\ \bibnamefont {O'Hara}},
  \bibinfo {author} {\bibfnamefont {T.~A.}\ \bibnamefont {Savard}}, \ and\
  \bibinfo {author} {\bibfnamefont {J.~E.}\ \bibnamefont {Thomas}},\
  }\href@noop {} {\bibfield  {journal} {\bibinfo  {journal} {Phys. Rev. A}\
  }\textbf {\bibinfo {volume} {58}},\ \bibinfo {pages} {3914} (\bibinfo {year}
  {1998})}\BibitemShut {NoStop}%
\bibitem [{\citenamefont {Grotz}\ \emph {et~al.}(2006)\citenamefont {Grotz},
  \citenamefont {Heaney},\ and\ \citenamefont {Strunz}}]{Grotz2006}%
  \BibitemOpen
  \bibfield  {author} {\bibinfo {author} {\bibfnamefont {T.}~\bibnamefont
  {Grotz}}, \bibinfo {author} {\bibfnamefont {L.}~\bibnamefont {Heaney}}, \
  and\ \bibinfo {author} {\bibfnamefont {W.~T.}\ \bibnamefont {Strunz}},\
  }\href@noop {} {\bibfield  {journal} {\bibinfo  {journal} {Phys. Rev. A}\
  }\textbf {\bibinfo {volume} {74}},\ \bibinfo {pages} {022102} (\bibinfo
  {year} {2006})}\BibitemShut {NoStop}%
\bibitem [{\citenamefont {D'Alessio}\ and\ \citenamefont
  {Rahmani}(2013)}]{D'Alessio2013}%
  \BibitemOpen
  \bibfield  {author} {\bibinfo {author} {\bibfnamefont {L.}~\bibnamefont
  {D'Alessio}}\ and\ \bibinfo {author} {\bibfnamefont {A.}~\bibnamefont
  {Rahmani}},\ }\href@noop {} {\bibfield  {journal} {\bibinfo  {journal} {Phys.
  Rev. B}\ }\textbf {\bibinfo {volume} {87}},\ \bibinfo {pages} {174301}
  (\bibinfo {year} {2013})}\BibitemShut {NoStop}%
\bibitem [{\citenamefont {Marino}\ and\ \citenamefont
  {Silva}(2012)}]{Marino2012}%
  \BibitemOpen
  \bibfield  {author} {\bibinfo {author} {\bibfnamefont {J.}~\bibnamefont
  {Marino}}\ and\ \bibinfo {author} {\bibfnamefont {A.}~\bibnamefont {Silva}},\
  }\href@noop {} {\bibfield  {journal} {\bibinfo  {journal} {Phys. Rev. B}\
  }\textbf {\bibinfo {volume} {86}},\ \bibinfo {pages} {060408} (\bibinfo
  {year} {2012})}\BibitemShut {NoStop}%
\bibitem [{\citenamefont {Marino}\ and\ \citenamefont
  {Silva}(2014)}]{Marino2014}%
  \BibitemOpen
  \bibfield  {author} {\bibinfo {author} {\bibfnamefont {J.}~\bibnamefont
  {Marino}}\ and\ \bibinfo {author} {\bibfnamefont {A.}~\bibnamefont {Silva}},\
  }\href@noop {} {\bibfield  {journal} {\bibinfo  {journal} {Phys. Rev. B}\
  }\textbf {\bibinfo {volume} {89}},\ \bibinfo {pages} {024303} (\bibinfo
  {year} {2014})}\BibitemShut {NoStop}%
\bibitem [{\citenamefont {Pichler}\ \emph {et~al.}(2012)\citenamefont
  {Pichler}, \citenamefont {Schachenmayer}, \citenamefont {Simon},
  \citenamefont {Zoller},\ and\ \citenamefont {Daley}}]{Pichler2012}%
  \BibitemOpen
  \bibfield  {author} {\bibinfo {author} {\bibfnamefont {H.}~\bibnamefont
  {Pichler}}, \bibinfo {author} {\bibfnamefont {J.}~\bibnamefont
  {Schachenmayer}}, \bibinfo {author} {\bibfnamefont {J.}~\bibnamefont
  {Simon}}, \bibinfo {author} {\bibfnamefont {P.}~\bibnamefont {Zoller}}, \
  and\ \bibinfo {author} {\bibfnamefont {A.~J.}\ \bibnamefont {Daley}},\
  }\href@noop {} {\bibfield  {journal} {\bibinfo  {journal} {Phys. Rev. A}\
  }\textbf {\bibinfo {volume} {86}},\ \bibinfo {pages} {051605} (\bibinfo
  {year} {2012})}\BibitemShut {NoStop}%
\bibitem [{\citenamefont {Pichler}\ \emph {et~al.}(2013)\citenamefont
  {Pichler}, \citenamefont {Schachenmayer}, \citenamefont {Daley},\ and\
  \citenamefont {Zoller}}]{Pichler2013}%
  \BibitemOpen
  \bibfield  {author} {\bibinfo {author} {\bibfnamefont {H.}~\bibnamefont
  {Pichler}}, \bibinfo {author} {\bibfnamefont {J.}~\bibnamefont
  {Schachenmayer}}, \bibinfo {author} {\bibfnamefont {A.~J.}\ \bibnamefont
  {Daley}}, \ and\ \bibinfo {author} {\bibfnamefont {P.}~\bibnamefont
  {Zoller}},\ }\href@noop {} {\bibfield  {journal} {\bibinfo  {journal} {Phys.
  Rev. A}\ }\textbf {\bibinfo {volume} {87}},\ \bibinfo {pages} {033606}
  (\bibinfo {year} {2013})}\BibitemShut {NoStop}%
\bibitem [{\citenamefont {Bunin}\ \emph {et~al.}(2011)\citenamefont {Bunin},
  \citenamefont {D'Alessio}, \citenamefont {Kafri},\ and\ \citenamefont
  {Polkovnikov}}]{Bunin2011}%
  \BibitemOpen
  \bibfield  {author} {\bibinfo {author} {\bibfnamefont {G.}~\bibnamefont
  {Bunin}}, \bibinfo {author} {\bibfnamefont {L.}~\bibnamefont {D'Alessio}},
  \bibinfo {author} {\bibfnamefont {Y.}~\bibnamefont {Kafri}}, \ and\ \bibinfo
  {author} {\bibfnamefont {A.}~\bibnamefont {Polkovnikov}},\ }\href@noop {}
  {\bibfield  {journal} {\bibinfo  {journal} {Nat. Phys.}\ }\textbf {\bibinfo
  {volume} {7}},\ \bibinfo {pages} {913} (\bibinfo {year} {2011})}\BibitemShut
  {NoStop}%
\bibitem [{\citenamefont {Prosen}(2008)}]{Prosen2008}%
  \BibitemOpen
  \bibfield  {author} {\bibinfo {author} {\bibfnamefont {T.}~\bibnamefont
  {Prosen}},\ }\href@noop {} {\bibfield  {journal} {\bibinfo  {journal} {New J.
  of Phys.}\ }\textbf {\bibinfo {volume} {10}},\ \bibinfo {pages} {043026}
  (\bibinfo {year} {2008})}\BibitemShut {NoStop}%
\bibitem [{\citenamefont {\ifmmode \check{Z}\else
  \v{Z}\fi{}nidari\ifmmode~\check{c}\else \v{c}\fi{}}(2010)}]{Znidaric2010}%
  \BibitemOpen
  \bibfield  {author} {\bibinfo {author} {\bibfnamefont {M.}~\bibnamefont
  {\ifmmode \check{Z}\else \v{Z}\fi{}nidari\ifmmode~\check{c}\else
  \v{c}\fi{}}},\ }\href@noop {} {\bibfield  {journal} {\bibinfo  {journal}
  {JSTAT}\ ,\ \bibinfo {pages} {L05002}} (\bibinfo {year} {2010})}\BibitemShut
  {NoStop}%
\bibitem [{\citenamefont {Eisler}(2011)}]{Eisler2011}%
  \BibitemOpen
  \bibfield  {author} {\bibinfo {author} {\bibfnamefont {V.}~\bibnamefont
  {Eisler}},\ }\href@noop {} {\bibfield  {journal} {\bibinfo  {journal}
  {JSTAT}\ ,\ \bibinfo {pages} {P06007}} (\bibinfo {year} {2011})}\BibitemShut
  {NoStop}%
\bibitem [{\citenamefont {\ifmmode \check{Z}\else
  \v{Z}\fi{}nidari\ifmmode~\check{c}\else \v{c}\fi{}}(2011)}]{Znidaric2011}%
  \BibitemOpen
  \bibfield  {author} {\bibinfo {author} {\bibfnamefont {M.}~\bibnamefont
  {\ifmmode \check{Z}\else \v{Z}\fi{}nidari\ifmmode~\check{c}\else
  \v{c}\fi{}}},\ }\href@noop {} {\bibfield  {journal} {\bibinfo  {journal}
  {Phys. Rev. E}\ }\textbf {\bibinfo {volume} {83}},\ \bibinfo {pages} {011108}
  (\bibinfo {year} {2011})}\BibitemShut {NoStop}%
\bibitem [{\citenamefont {Hartmann}\ \emph {et~al.}(2009)\citenamefont
  {Hartmann}, \citenamefont {Prior}, \citenamefont {Clark},\ and\ \citenamefont
  {Plenio}}]{Hartmann2009}%
  \BibitemOpen
  \bibfield  {author} {\bibinfo {author} {\bibfnamefont {M.~J.}\ \bibnamefont
  {Hartmann}}, \bibinfo {author} {\bibfnamefont {J.}~\bibnamefont {Prior}},
  \bibinfo {author} {\bibfnamefont {S.~R.}\ \bibnamefont {Clark}}, \ and\
  \bibinfo {author} {\bibfnamefont {M.~B.}\ \bibnamefont {Plenio}},\
  }\href@noop {} {\bibfield  {journal} {\bibinfo  {journal} {Phys. Rev. Lett.}\
  }\textbf {\bibinfo {volume} {102}},\ \bibinfo {pages} {057202} (\bibinfo
  {year} {2009})}\BibitemShut {NoStop}%
\bibitem [{\citenamefont {Atala}\ \emph {et~al.}(2013)\citenamefont {Atala},
  \citenamefont {Aidelsburger}, \citenamefont {Barreiro}, \citenamefont
  {Abanin}, \citenamefont {Kitagawa}, \citenamefont {Demler},\ and\
  \citenamefont {Bloch}}]{Atala2013}%
  \BibitemOpen
  \bibfield  {author} {\bibinfo {author} {\bibfnamefont {M.}~\bibnamefont
  {Atala}}, \bibinfo {author} {\bibfnamefont {M.}~\bibnamefont {Aidelsburger}},
  \bibinfo {author} {\bibfnamefont {J.~T.}\ \bibnamefont {Barreiro}}, \bibinfo
  {author} {\bibfnamefont {D.}~\bibnamefont {Abanin}}, \bibinfo {author}
  {\bibfnamefont {T.}~\bibnamefont {Kitagawa}}, \bibinfo {author}
  {\bibfnamefont {E.}~\bibnamefont {Demler}}, \ and\ \bibinfo {author}
  {\bibfnamefont {I.}~\bibnamefont {Bloch}},\ }\href@noop {} {\bibfield
  {journal} {\bibinfo  {journal} {Nature Phys.}\ }\textbf {\bibinfo {volume}
  {9}},\ \bibinfo {pages} {795} (\bibinfo {year} {2013})}\BibitemShut {NoStop}%
\bibitem [{\citenamefont {Breuer}\ and\ \citenamefont
  {Petruccione}(2002)}]{Breuer2002}%
  \BibitemOpen
  \bibfield  {author} {\bibinfo {author} {\bibfnamefont {H.-P.}\ \bibnamefont
  {Breuer}}\ and\ \bibinfo {author} {\bibfnamefont {F.}~\bibnamefont
  {Petruccione}},\ }\href@noop {} {\emph {\bibinfo {title} {The theory of open
  quantum systems}}}\ (\bibinfo  {publisher} {Oxford University Press},\
  \bibinfo {year} {2002})\BibitemShut {NoStop}%
\bibitem [{\citenamefont {Barchielli}\ and\ \citenamefont
  {Gregoratti}(2009)}]{Barchielli2009}%
  \BibitemOpen
  \bibfield  {author} {\bibinfo {author} {\bibfnamefont {A.}~\bibnamefont
  {Barchielli}}\ and\ \bibinfo {author} {\bibfnamefont {M.}~\bibnamefont
  {Gregoratti}},\ }\href@noop {} {\emph {\bibinfo {title} {Quantum Trajectories
  and Measurements in Continuous Time}}}\ (\bibinfo  {publisher}
  {Springer-Verlag, Berlin},\ \bibinfo {year} {2009})\BibitemShut {NoStop}%
\bibitem [{\citenamefont {Caves}\ and\ \citenamefont
  {Milburn}(1987)}]{Caves1987}%
  \BibitemOpen
  \bibfield  {author} {\bibinfo {author} {\bibfnamefont {C.~M.}\ \bibnamefont
  {Caves}}\ and\ \bibinfo {author} {\bibfnamefont {G.~J.}\ \bibnamefont
  {Milburn}},\ }\href@noop {} {\bibfield  {journal} {\bibinfo  {journal} {Phys.
  Rev. A}\ }\textbf {\bibinfo {volume} {36}},\ \bibinfo {pages} {5543}
  (\bibinfo {year} {1987})}\BibitemShut {NoStop}%
\bibitem [{\citenamefont {Di\'osi}\ and\ \citenamefont
  {Kiefer}(2000)}]{Diosi2000}%
  \BibitemOpen
  \bibfield  {author} {\bibinfo {author} {\bibfnamefont {L.}~\bibnamefont
  {Di\'osi}}\ and\ \bibinfo {author} {\bibfnamefont {C.}~\bibnamefont
  {Kiefer}},\ }\href@noop {} {\bibfield  {journal} {\bibinfo  {journal} {Phys.
  Rev. Lett.}\ }\textbf {\bibinfo {volume} {85}},\ \bibinfo {pages} {3552}
  (\bibinfo {year} {2000})}\BibitemShut {NoStop}%
\bibitem [{\citenamefont {Dalla~Torre}\ \emph {et~al.}(2010)\citenamefont
  {Dalla~Torre}, \citenamefont {Demler}, \citenamefont {Giamarchi},\ and\
  \citenamefont {Altman}}]{DallaTorre2010}%
  \BibitemOpen
  \bibfield  {author} {\bibinfo {author} {\bibfnamefont {E.~G.}\ \bibnamefont
  {Dalla~Torre}}, \bibinfo {author} {\bibfnamefont {E.}~\bibnamefont {Demler}},
  \bibinfo {author} {\bibfnamefont {T.}~\bibnamefont {Giamarchi}}, \ and\
  \bibinfo {author} {\bibfnamefont {E.}~\bibnamefont {Altman}},\ }\href@noop {}
  {\bibfield  {journal} {\bibinfo  {journal} {Nat. Phys.}\ }\textbf {\bibinfo
  {volume} {6}},\ \bibinfo {pages} {806} (\bibinfo {year} {2010})}\BibitemShut
  {NoStop}%
\bibitem [{\citenamefont {Wilson}\ \emph {et~al.}(2012)\citenamefont {Wilson},
  \citenamefont {Fregoso},\ and\ \citenamefont {Galitski}}]{Wilson2012}%
  \BibitemOpen
  \bibfield  {author} {\bibinfo {author} {\bibfnamefont {J.~H.}\ \bibnamefont
  {Wilson}}, \bibinfo {author} {\bibfnamefont {B.~M.}\ \bibnamefont {Fregoso}},
  \ and\ \bibinfo {author} {\bibfnamefont {V.~M.}\ \bibnamefont {Galitski}},\
  }\href@noop {} {\bibfield  {journal} {\bibinfo  {journal} {Phys. Rev. B}\
  }\textbf {\bibinfo {volume} {85}},\ \bibinfo {pages} {174304} (\bibinfo
  {year} {2012})}\BibitemShut {NoStop}%
\bibitem [{\citenamefont {Klich}(2003)}]{Klich2003}%
  \BibitemOpen
  \bibfield  {author} {\bibinfo {author} {\bibfnamefont {I.}~\bibnamefont
  {Klich}},\ }\href@noop {} {\emph {\bibinfo {title} {Quantum Noise in
  Mesoscopic Systems}}},\ edited by\ \bibinfo {editor} {\bibfnamefont {Y.~V.}\
  \bibnamefont {Nazarov}}\ (\bibinfo  {publisher} {Kluwer},\ \bibinfo {year}
  {2003})\BibitemShut {NoStop}%
\bibitem [{Kli()}]{Klich2003'}%
  \BibitemOpen
  \href@noop {} {}\Eprint {http://arxiv.org/abs/arXiv:cond-mat/0209642}
  {arXiv:cond-mat/0209642} \BibitemShut {NoStop}%
\bibitem [{\citenamefont {Jackiw}\ and\ \citenamefont
  {Rebbi}(1976)}]{Jackiw1976}%
  \BibitemOpen
  \bibfield  {author} {\bibinfo {author} {\bibfnamefont {R.}~\bibnamefont
  {Jackiw}}\ and\ \bibinfo {author} {\bibfnamefont {C.}~\bibnamefont {Rebbi}},\
  }\href@noop {} {\bibfield  {journal} {\bibinfo  {journal} {Phys. Rev. D}\
  }\textbf {\bibinfo {volume} {13}},\ \bibinfo {pages} {3398} (\bibinfo {year}
  {1976})}\BibitemShut {NoStop}%
\bibitem [{\citenamefont {Su}\ \emph {et~al.}(1979)\citenamefont {Su},
  \citenamefont {Schrieffer},\ and\ \citenamefont {Heeger}}]{Su1979}%
  \BibitemOpen
  \bibfield  {author} {\bibinfo {author} {\bibfnamefont {W.~P.}\ \bibnamefont
  {Su}}, \bibinfo {author} {\bibfnamefont {J.~R.}\ \bibnamefont {Schrieffer}},
  \ and\ \bibinfo {author} {\bibfnamefont {A.~J.}\ \bibnamefont {Heeger}},\
  }\href@noop {} {\bibfield  {journal} {\bibinfo  {journal} {Phys. Rev. Lett.}\
  }\textbf {\bibinfo {volume} {42}},\ \bibinfo {pages} {1698} (\bibinfo {year}
  {1979})}\BibitemShut {NoStop}%
\bibitem [{\citenamefont {Hou}\ \emph {et~al.}(2007)\citenamefont {Hou},
  \citenamefont {Chamon},\ and\ \citenamefont {Mudry}}]{Hou2007}%
  \BibitemOpen
  \bibfield  {author} {\bibinfo {author} {\bibfnamefont {C.-Y.}\ \bibnamefont
  {Hou}}, \bibinfo {author} {\bibfnamefont {C.}~\bibnamefont {Chamon}}, \ and\
  \bibinfo {author} {\bibfnamefont {C.}~\bibnamefont {Mudry}},\ }\href@noop {}
  {\bibfield  {journal} {\bibinfo  {journal} {Phys. Rev. Lett.}\ }\textbf
  {\bibinfo {volume} {98}},\ \bibinfo {pages} {186809} (\bibinfo {year}
  {2007})}\BibitemShut {NoStop}%
\bibitem [{\citenamefont {Seradjeh}\ and\ \citenamefont
  {Franz}(2008)}]{Seradjeh2008}%
  \BibitemOpen
  \bibfield  {author} {\bibinfo {author} {\bibfnamefont {B.}~\bibnamefont
  {Seradjeh}}\ and\ \bibinfo {author} {\bibfnamefont {M.}~\bibnamefont
  {Franz}},\ }\href@noop {} {\bibfield  {journal} {\bibinfo  {journal} {Phys.
  Rev. Lett.}\ }\textbf {\bibinfo {volume} {101}},\ \bibinfo {pages} {146401}
  (\bibinfo {year} {2008})}\BibitemShut {NoStop}%
\bibitem [{\citenamefont {Rahmani}\ \emph
  {et~al.}(2013{\natexlab{a}})\citenamefont {Rahmani}, \citenamefont {Muniz},\
  and\ \citenamefont {Martin}}]{Rahmani2013b}%
  \BibitemOpen
  \bibfield  {author} {\bibinfo {author} {\bibfnamefont {A.}~\bibnamefont
  {Rahmani}}, \bibinfo {author} {\bibfnamefont {R.~A.}\ \bibnamefont {Muniz}},
  \ and\ \bibinfo {author} {\bibfnamefont {I.}~\bibnamefont {Martin}},\
  }\href@noop {} {\bibfield  {journal} {\bibinfo  {journal} {Phys. Rev. X}\
  }\textbf {\bibinfo {volume} {3}},\ \bibinfo {pages} {031008} (\bibinfo {year}
  {2013}{\natexlab{a}})}\BibitemShut {NoStop}%
\bibitem [{\citenamefont {Chen}\ \emph {et~al.}(2010)\citenamefont {Chen},
  \citenamefont {Ruschhaupt}, \citenamefont {Schmidt}, \citenamefont {del
  Campo}, \citenamefont {Gu\'ery-Odelin},\ and\ \citenamefont
  {Muga}}]{Chen2010}%
  \BibitemOpen
  \bibfield  {author} {\bibinfo {author} {\bibfnamefont {X.}~\bibnamefont
  {Chen}}, \bibinfo {author} {\bibfnamefont {A.}~\bibnamefont {Ruschhaupt}},
  \bibinfo {author} {\bibfnamefont {S.}~\bibnamefont {Schmidt}}, \bibinfo
  {author} {\bibfnamefont {A.}~\bibnamefont {del Campo}}, \bibinfo {author}
  {\bibfnamefont {D.}~\bibnamefont {Gu\'ery-Odelin}}, \ and\ \bibinfo {author}
  {\bibfnamefont {J.~G.}\ \bibnamefont {Muga}},\ }\href@noop {} {\bibfield
  {journal} {\bibinfo  {journal} {Phys. Rev. Lett.}\ }\textbf {\bibinfo
  {volume} {104}},\ \bibinfo {pages} {063002} (\bibinfo {year}
  {2010})}\BibitemShut {NoStop}%
\bibitem [{\citenamefont {Doria}\ \emph {et~al.}(2011)\citenamefont {Doria},
  \citenamefont {Calarco},\ and\ \citenamefont {Montangero}}]{Doria2011}%
  \BibitemOpen
  \bibfield  {author} {\bibinfo {author} {\bibfnamefont {P.}~\bibnamefont
  {Doria}}, \bibinfo {author} {\bibfnamefont {T.}~\bibnamefont {Calarco}}, \
  and\ \bibinfo {author} {\bibfnamefont {S.}~\bibnamefont {Montangero}},\
  }\href@noop {} {\bibfield  {journal} {\bibinfo  {journal} {Phys. Rev. Lett.}\
  }\textbf {\bibinfo {volume} {106}},\ \bibinfo {pages} {190501} (\bibinfo
  {year} {2011})}\BibitemShut {NoStop}%
\bibitem [{\citenamefont {Rahmani}\ and\ \citenamefont
  {Chamon}(2011)}]{Rahmani2011}%
  \BibitemOpen
  \bibfield  {author} {\bibinfo {author} {\bibfnamefont {A.}~\bibnamefont
  {Rahmani}}\ and\ \bibinfo {author} {\bibfnamefont {C.}~\bibnamefont
  {Chamon}},\ }\href@noop {} {\bibfield  {journal} {\bibinfo  {journal} {Phys.
  Rev. Lett.}\ }\textbf {\bibinfo {volume} {107}},\ \bibinfo {pages} {016402}
  (\bibinfo {year} {2011})}\BibitemShut {NoStop}%
\bibitem [{\citenamefont {Choi}\ \emph {et~al.}(2012)\citenamefont {Choi},
  \citenamefont {Onofrio},\ and\ \citenamefont {Sundaram}}]{Choi2012}%
  \BibitemOpen
  \bibfield  {author} {\bibinfo {author} {\bibfnamefont {S.}~\bibnamefont
  {Choi}}, \bibinfo {author} {\bibfnamefont {R.}~\bibnamefont {Onofrio}}, \
  and\ \bibinfo {author} {\bibfnamefont {B.}~\bibnamefont {Sundaram}},\
  }\href@noop {} {\bibfield  {journal} {\bibinfo  {journal} {Phys. Rev. A}\
  }\textbf {\bibinfo {volume} {86}},\ \bibinfo {pages} {043436} (\bibinfo
  {year} {2012})}\BibitemShut {NoStop}%
\bibitem [{\citenamefont {Ruschhaupt}\ \emph {et~al.}(2012)\citenamefont
  {Ruschhaupt}, \citenamefont {Chen}, \citenamefont {Alonso},\ and\
  \citenamefont {Muga}}]{Ruschhaupt2012}%
  \BibitemOpen
  \bibfield  {author} {\bibinfo {author} {\bibfnamefont {A.}~\bibnamefont
  {Ruschhaupt}}, \bibinfo {author} {\bibfnamefont {X.}~\bibnamefont {Chen}},
  \bibinfo {author} {\bibfnamefont {D.}~\bibnamefont {Alonso}}, \ and\ \bibinfo
  {author} {\bibfnamefont {J.~G.}\ \bibnamefont {Muga}},\ }\href@noop {}
  {\bibfield  {journal} {\bibinfo  {journal} {New J. Phys.}\ }\textbf {\bibinfo
  {volume} {14}},\ \bibinfo {pages} {093040} (\bibinfo {year}
  {2012})}\BibitemShut {NoStop}%
\bibitem [{\citenamefont {Rahmani}\ \emph
  {et~al.}(2013{\natexlab{b}})\citenamefont {Rahmani}, \citenamefont
  {Kitagawa}, \citenamefont {Demler},\ and\ \citenamefont
  {Chamon}}]{Rahmani2013a}%
  \BibitemOpen
  \bibfield  {author} {\bibinfo {author} {\bibfnamefont {A.}~\bibnamefont
  {Rahmani}}, \bibinfo {author} {\bibfnamefont {T.}~\bibnamefont {Kitagawa}},
  \bibinfo {author} {\bibfnamefont {E.}~\bibnamefont {Demler}}, \ and\ \bibinfo
  {author} {\bibfnamefont {C.}~\bibnamefont {Chamon}},\ }\href@noop {}
  {\bibfield  {journal} {\bibinfo  {journal} {Phys. Rev. A}\ }\textbf {\bibinfo
  {volume} {87}},\ \bibinfo {pages} {043607} (\bibinfo {year}
  {2013}{\natexlab{b}})}\BibitemShut {NoStop}%
\end{thebibliography}%

\end{document}